\documentclass[10pt,bezier]{article}
\usepackage{amsmath,amssymb,amsfonts, euscript}
\newcommand{\R}{I\!\!R}
\newcommand{\p}{\partial}
\newcommand{\n}{\nabla}
\newcommand{\X}{\frak{X}}
\newcommand{\hn}{\hat \nabla}
\newcommand{\vn}{\vec \nabla}
\newcommand{\bn}{\bar \nabla}
\newcommand{\z}{{\bf Z}_2}
\newcommand{\lon}{\longrightarrow}
\newcommand{\te}{\textstyle}
\newtheorem{thm}{Theorem}[section]

\newtheorem{defn}[thm]{Definition}

\begin{document}
\noindent {\bf\Large  Geometrization  of Mass in General Relativity}\\

\noindent \\

\noindent  {\bf  Nasser  boroojerdian}\footnote{N. boroojerdian\\
\indent  Dep. of Math. \& Comp., Amirkabir University of Technology, Tehran, Iran \\
\indent  E-mail: broojerd@aut.ac.ir}\\

\noindent \\

\noindent {\bf  Abstract:} In this paper we will extend the notion
of tangent bundle to a $\z$ graded tangent bundle. This graded
bundle has a Lie algebroid structure and we can develop notions
semi-riemannian metrics, Levi-civita connection, and curvature, on
it. In case of space-times manifolds, even part of the tangent
bundle is related to space and time structures(gravity) and odd
part is related to mass distribution in space-time. In this
structure, mass becomes part of the geometry, and Einstein field
equation can be
reconstructed in a new simpler form. The new field equation is purely geometric.   \\


\section{Graded tangent bundles}
In general relativity, gravity can be formulated by Lorentzian
metrics on the ordinary tangent bundle of a space-time.  Science
we can extend ordinary tangent bundle of a manifold to a larger
bundle, so we obtain additional degree of freedom to describe
more concepts.

Tangent vectors to a manifold $M$, can be identified by
point-derivations of the algebra $C^\infty(M)$. If we replace the
algebra $C^\infty(M)$ by some other related algebra, we may find
some new tangent vectors. Considering the $\z$ graded two
dimensional algebra $\R\oplus\R$, this algebra is the Clifford
algebra of $\R$ furnished with its canonical positive definite
inner product. The unit of this algebra is $(1,1)$ which is
denoted by {\bf 1}. Even part of the algebra $\R\oplus\R$ is
generated by {\bf 1}. Odd part of this algebra is generated by
$\tau=(1,-1)$. So, every elements of $\R\oplus\R$ has the form
$\lambda{\bf 1}+\mu\tau$, in fact
\[ (a,b)={a+b\over 2}{\bf 1}+{a-b\over 2}\tau \]
For the sake of simplicity, denote $\lambda{\bf 1}$ by $\lambda$.
Denote the set of all $\R\oplus\R$ valued smooth function on $M$,
by $\hat C^\infty(M)$. This set, by pointwise addition and
multiplication is a $\z$ graded algebra, and its elements have a
form of $f+g\tau$ in which $f,g\in C^\infty(M)$. Even subalgebra
of $\hat C^\infty(M)$ is $C^\infty(M)$, and its odd functions have
a form of $g\tau$ for some $g\in C^\infty(M)$.  Even and odd
elements of a $\z$ graded algebra are called homogeneous and
their parity is defined as follows
\[ |a|=\left\{ \begin{array}{cc} 0& a\ {\rm is\ even.}\\ 1& a\ {\rm
is\ odd.} \end{array} \right. \] Derivations of $C^\infty(M)$
identify ordinary vector fields on $M$. Now, we are going to find
graded derivations of $\hat C^\infty(M)$.

\begin{defn}
A linear map $D:\hat C^\infty(M)\longrightarrow\hat C^\infty(M)$
is called an even derivation iff even and odd subspaces of $\hat
C^\infty(M)$ are invariant under $D$ and for \mbox{$\hat f,\hat
g\in \hat C^\infty(M)$}, we have
\[ D(\hat f\hat g)=D(\hat f)\hat g+\hat f D(\hat g) \]
\end{defn}
\begin{defn}
A linear map $D:\hat C^\infty(M)\longrightarrow\hat C^\infty(M)$
is called an odd derivation iff $D$ changes parity of homogeneous
elements and for homogeneous elements \mbox{$\hat f,\hat g\in \hat
C^\infty(M)$}, we have
\[ D(\hat f\hat g)=D(\hat f)\hat g+(-1)^{|\hat f|}\hat f D(\hat g) \]
\end{defn}
\begin{thm}
Any vector field $X\in\X M$ determine an even derivation on $\hat
C^\infty(M)$ by $X(f+g\tau)=X(f)+X(g)\tau$ and every even
derivation on $\hat C^\infty(M)$ can be obtained by this way.
\end{thm}
{\bf Proof:} Clearly the operation of $X$ on $\hat C^\infty(M)$
defined as above, is an even derivation. Conversely, let $D$ be an
even derivation. Restriction of $D$ to $C^\infty(M)$ is an
ordinary derivation on $C^\infty(M)$ and determines some vector
field $X\in\X M$ such that for \mbox{$f\in C^\infty(M)$} we have
$D(f)=X(f)$. We can deduce $D(\tau)=0$.
\[ 0=D(1)=D(\tau\tau)=2\tau D(\tau)\Rightarrow D(\tau)=0 \]
It is easy to see that $D(f+g\tau)=X(f)+X(g)\tau\ .\quad\bullet$
\begin{thm}
Any function $h\in C^\infty(M)$ determine an odd derivation on
$\hat C^\infty(M)$ by $D_h(f+g\tau)=gh$ and every odd derivation
on $\hat C^\infty(M)$ can be obtained by this way.
\end{thm}
{\bf Proof:} Clearly  $D_h$ is an odd derivation. Conversely,
suppose that $D$ is an odd derivation. By commutativity of $\hat
C^\infty(M)$, we can infer that for any function $f\in
C^\infty(M)$ we have $D(f)=0$.
\[
D(f)\tau+fD(\tau)=D(f\tau)=D(\tau f)=D(\tau)f-\tau D(f)
\Rightarrow 2D(f)\tau =0 \Rightarrow D(f)=0
\]
$D(\tau)$ is an even element of $\hat C^\infty(M)$, so for some
$h\in C^\infty(M)$ we have $D(\tau)=h$. It is easy to show
 that $D=D_h .\quad\bullet$

These theorems show that the space of graded derivations of the
graded algebra $\hat C^\infty(M)$ is $\X(M)\oplus C^\infty(M)$. By
pointwise addition and multiplication, this space is a $\z$
graded module on $C^\infty(M)$, and we denote it by $\hat\X(M)$.
Even part of $\hat\X(M)$ is $\X(M)$ and its odd part is
$C^\infty(M)$. For constant function {\bf 1}, denote the odd
derivation $D_{\bf\te 1}$ by $\xi$, so $\xi(f+g\tau)=g$. Every odd
derivation on $\hat C^\infty(M)$ is of the form $h\xi$ for some
$h\in C^\infty(M)$, and we call them odd vector fields on $M$.
Even vector fields, are ordinary vector fields on $M$. We can
construct the bundle $\hat TM$ as follows
\[ \hat TM=\bigcup_{p\in M}T_pM\oplus\R \]
$\hat\X(M)$ is the space of sections of $\hat TM$, so we call
$\hat TM$ as the graded tangent bundle of $M$. $\hat TM$ is a $\z$
graded bundle and is a natural graded extension of the ordinary
tangent bundle. Henceforth, we use symbols $\hat X,\hat Y,\hat
Z,...$ for arbitrary sections of $\hat TM$ and symbols $X,Y,Z,...$
for ordinary (or even) vector fields on $M$. Lie bracket of two
graded derivation $D$ and $D'$, is a graded derivation, defined as
follows:
\[ [D\ ,\ D']=D\circ D'-(-1)^{|D||D'|}D'\circ D \]
This definition implies that lie bracket of even vector fields is
the ordinary lie bracket of vector fields, and lie bracket of even
and odd vector fields is as follows:
\begin{eqnarray*}
{[}X\ ,\  h\xi]&=&X(h)\xi\\
{[} f\xi\ ,\ g\xi]&=&0
\end{eqnarray*}
This lie bracket, make $\hat C^\infty(M)$ into a super lie
algebra. Since, lie bracket of odd vector fields are zero, $\hat
C^\infty(M)$ is an ordinary lie algebra too. This lie algebra
structure of $\hat C^\infty(M)$ and the anchor map $\rho :\hat
TM\lon TM,\ \rho(X+h\xi)=X$, convert $\hat TM$ into a lie
algebroid. So, we can use properties of algebroid structures for
$\hat TM$.

\section{Graded connections on graded tangent bundles}
Due to the theory of connections on lie algebroids, a connection
on $\hat TM$ is an operator
\mbox{$\hn:\hat\X(M)\times\hat\X(M)\lon\hat\X(M)$}
that satisfies the following relations.\\
 For $\hat X,\hat Y\in\hat\X(M)$ and $f\in C^\infty(M):$
\[\begin{array}{rcl}
{\hn}_{f\hat X}\hat Y&=&f{\hn}_{\hat X}\hat Y\\
{\hn}_{\hat X}f\hat Y&=&\rho(\hat X)(f)\hat Y+f{\hn}_{\hat X}\hat
Y
\end{array}\]
Note that here, we have $\rho(\hat X)(f)=\hat X(f)$, and we can
rewrite the second equation in a more simple and natural form. If
a connection $\hn$ on $\hat TM$, respects parity of vector fields
such that for homogeneous vector fields $\hat X,\hat Y$,
$\hn_{\hat X}\hat Y$ be homogeneous and \mbox{$|\hn_{\hat X}\hat
Y|=|\hat X|+|\hat Y|$}, then we call it a graded connection.
\begin{thm}
For any graded connection $\hn$ on $\hat TM$, there exist a
unique connection $\n$ on $M$ and two 1-form $\alpha,\alpha'$ and
a vector field $X_0$ on $M$ such that for $X,Y\in\X(M)$ and
$h,k\in C^\infty(M)$:
\begin{eqnarray}
\hn_XY&=&\nabla_XY\\
\hn_{h\xi}X&=&h\alpha(X)\xi\\
\hn_Xh\xi&=&h\alpha'(X)\xi+X(h)\xi\\
\hn_{h\xi}k\xi&=&hkX_0
\end{eqnarray}
\end{thm}
{\bf Proof:} Restriction of $\hn$ to even vector fields is an
ordinary connection $\n$ on $M$. $\hn_\xi X$ and $\hn_X\xi$ are
odd vector fields, so we can define $\alpha$ and $\alpha'$ by
$\alpha(X)\xi=\hn_\xi X$, $\alpha'(X)\xi=\hn_X\xi$. Properties of
connections and odd vector fields imply that $\alpha$ and
$\alpha'$ are 1-forms. Set $X_0=\hn_\xi\xi$. Straight
computations yield equations (1)-(4). $\bullet$

Conversely, any graded connection on $\hat TM$ is obtained by
equations(1)-(4) for some connection $\n$ and 1-forms $\alpha$
and $\alpha'$ and some vector field $X_0$ on $M$. Torsion of
$\hn$ is defined as follows.
\[ \hat T(\hat X,\ \hat Y)=\hn_{\hat X}\hat Y-\hn_{\hat Y}\hat X-[\hat X,\ \hat Y]\]
$\hn$ is torsion free iff $\n$ is torsion free and
$\alpha=\alpha'$. So, torsion free graded connections on $\hat
TM$ are obtained by triplet $(\n ,\alpha, X_0)$ in which $\n$ is a
torsion free connection and $\alpha$ is a 1-form and $X_0$ is a
vector field on $M$.

\section{Semi-Riemannian metrics on graded tangent bundles}
\begin{defn}
Any semi-riemannian metric on the vector bundle $\hat TM$ is
called a graded metric on $\hat TM$ iff even vectors are
orthogonal to odd vectors.
\end{defn}
Any semi-riemannian metrics on $\hat TM$ determines a compatible
and torsion free connection $\hn$ on $\hat TM$ by koszul formula
[2], and is called Levi-Civita connection of the meter.
\[\begin{array}{rl}
2<\hn_{\hat X}\hat Y,\hat Z>=&\hat X<\hat Y,\hat Z>+\hat Y<\hat Z,\hat X>-\hat Z<\hat X,\hat Y>\\
 &<[\hat X,\hat Y],\hat Z>-<[\hat Y,\hat Z],\hat X>+<[\hat Z,\hat X],\hat Y>
\end{array}\]
\begin{thm}
Levi-Civita connection of a graded metric on $\hat TM$ is a
graded connection.
\end{thm}
{\bf Proof:} If sum of the parity of homogeneous vector fields
$\hat X,\ \hat Y,\ \hat Z$ be odd, then Koszul formula shows that
$\hn_{\hat X}{\hat Y}$ is orthogonal to $\hat Z$. So, $\hn_{\hat
X}{\hat Y}$ is homogeneus and its parity is sum of the parity of
$\hat X$ and $\hat Y.\ \bullet$

Let $\hat g$ be a graded metric on $\hat TM$, then its restriction
to $TM$ is a semi-riemannian metric on $M$ and $h=\hat g(\xi,\xi)$
is a smooth nonzero function on $M$. Conversely, every
semi-riemannian metric $g$ on $M$ and smooth nonzero function $h$
on $M$, determine a graded metric on $\hat TM$ as follows:
\[ <X+f\xi\ ,\ Y+k\xi>=g(X,Y)+fkh. \]
Without lose of generality, we consider the case $\hat g(\xi,\xi)$
is positive and is of the form $\hat g(\xi,\xi)=e^{2\theta}$. So,
graded metrics on $M$ are determned by pairs $(g,\theta)$ in
which $g$ is a semi-riemanian metric on $M$ and $\theta\in
C^\infty(M)$ such that $<X,Y>=g(X,Y)$ and
$<\xi,\xi>=e^{2\theta}.$\\
The gradient of a smooth function $f$ on a semi-riemannian
manifold $(M,g)$ is denoted by $\vn f$ and defined by $g(\vn
f,X)=df(X)=X(f)$. As we have already shown, any torsion free
graded connection $\hn$ on $\hat TM$ is determined by a triplet
$(\n,\alpha,X_0)$, in which $\n$ is a torsion free connection and
$\alpha$ is a 1-form and $X_0$ is a vector field on $M$.
\begin{thm}
If $\hat g =(g,\theta)$ is a graded semi-riemannian metric on
$\hat TM$, then its Levi-Civita connection is determined by the
triplet $(\n,d\theta,-e^{2\theta}\vn\theta)$ in which $\n$ is
Levi-Civita connection of $g$.
\end{thm}
{\bf Proof:} Let Levi-Civita connection of $\hat g$ be determined
by $(\n,\alpha,X_0)$. Applying Koszul formula to even vector
fields implies that $\n$ is the Levi-Civita connection of $g$.
Now,
\begin{eqnarray*}
X(e^{2\theta})&=&X<\xi,\xi>=2<\n_X\xi,\xi>=2<\alpha(X)\xi,\xi>\\
&=&2\alpha(X)<\xi,\xi>=2\alpha(X)e^{2\theta}
\end{eqnarray*}
 But $X(e^{2\theta})=2e^{2\theta}X(\theta)$, so
$\alpha(X)=X(\theta)$. Consequently, $\alpha=d\theta$. Following
computations show that $X_0=-e^{2\theta}\vn\theta$.
\[\begin{array}{rl}
0&=\xi<Y,\xi>=<\n_\xi Y,\xi>+<Y,\n_\xi\xi>\\
 &=<\alpha(Y)\xi,\xi>+<Y,X_0>=\alpha(Y)<\xi,\xi>+<Y,X_0>\\
 &=e^{2\theta}d\theta(Y)+<Y,X_0>\\
 \Rightarrow & <X_0,Y>=<-e^{2\theta}\vn\theta\ ,Y> \ \ \Rightarrow \ \  X_0=-e^{2\theta}\vn\theta\quad\bullet
\end{array}\]
Explicitly, $\hn$ is determined as follows.
\begin{eqnarray}
\hn_XY &=& \n_XY\\
\hn_\xi X=\hn_X\xi&=&d\theta(X)\xi\\
\hn_\xi\xi &=& -e^{2\theta}\vn\theta
\end{eqnarray}

\section{Curvature tensors of graded metrics}
In this section $\hat g =(g,\theta)$ is a graded metric on $\hat
TM$ and $\hn =(\n,\alpha,X_0)$ is its Levi-Civita connection.
Curvature tensors of $\hn$ as a connection in algebroid
structures are defined. Curvature and Ricci curvature tensors of
$\hat g$ and $g$ are denoted respectively by $\hat {\rm
R},\widehat{\rm Ric}, {\rm R}, {\rm Ric}$.
\begin{thm}
Curvature tensor $\hat R$ respect parity of homogeneous vector
fields and satisfies the following relations.
\begin{eqnarray}
\hat {\rm R}(X,Y)(Z) &=& {\rm R}(X,Y)(Z)\\
\hat {\rm R}(X,Y)(\xi)&=&0\\
\hat {\rm R}(X,\xi)(Y)&=&(\nabla_X\alpha)(Y)\xi+\alpha(X)\alpha(Y)\xi\\
\hat {\rm R}(X,\xi)(\xi)&=&\nabla_XX_0-\alpha(X)X_0
\end{eqnarray}
\end{thm}
{\bf Proof:} Formula of the curvature computation shows that the
curvature tensor respect parity of vector fields.By straight
forward computations we find equations (8)-(11) hold. For example
to prove (9), assume $[X,Y]=0$, so
\[\begin{array}{rcl}
\hat {\rm R}(X,Y)(\xi)&=&\hn_X\hn_Y\xi-\hn_Y\hn_X\xi=\hn_X\alpha(Y)\xi-\hn_Y\alpha(X)\xi\\
 &=&X(\alpha(Y))\xi+\alpha(Y)\hn_X\xi-Y(\alpha(X))\xi-\alpha(X)\hn_Y\xi\\
 &=&d\alpha(X,Y)\xi+\alpha(Y)\alpha(X)\xi-\alpha(X)\alpha(Y)\xi=0
\end{array}\]
Note that $d\alpha=0$ because $\alpha=d\theta.\ \bullet$

Scalar part of the $\hat {\rm R}(X,\xi)(Y)$ is a tensor with
respect to $X$ and $Y$ and it is convenient to have some name for
it. Set

\[ \tilde T(X,Y)=(\nabla_X\alpha)(Y)+\alpha(X)\alpha(Y) \]
Since $\alpha$ is closed and $\n$ is torsion free, they imply that
$(\nabla_X\alpha)(Y)$ is symmetric with respect to $X$ and $Y$.
\begin{eqnarray*}
(\n_X\alpha)(Y)&=&(\n_X\alpha)(Y)-(\n_Y\alpha)(X)+(\n_Y\alpha)(X)\\
 &=&d\alpha(X,Y)+(\n_Y\alpha)(X)=(\n_Y\alpha)(X)
\end{eqnarray*}
Consequently, $\tilde T$ is a symmetric tensor. For a smooth
function $f$ on a semi-riemannian manifold $(M,g)$, its Hessian is
a 2-covariant symmetric tensor denoted by ${\rm Hes}(f)$ and is
defied as follows.
\[{\rm Hes}(f)(X,Y)=(\nabla_Xdf)(Y)\]
So, the symmetric tensor $\tilde T$ can be written as follows.
\begin{equation}
 \tilde T={\rm Hes}(\theta)+d\theta\otimes d\theta
 \end{equation}
Laplacian of a smooth function $f$ is also a smooth function
denoted by $\triangle (f)$ and is defined by $\triangle (f)={\rm
div}(df)={\rm tr}(\rm Hes(f))$.\\
 Since, ${\rm tr}(d\theta\otimes
d\theta)=g(\vn\theta,\vn\theta)=|\vn\theta|^2$, we find
\[ {\rm tr}(\tilde T)=\triangle (\theta)+|\vn \theta|^2 \]
\begin{thm}
Ricci curvature of $\hat g$ satisfies the following relations.
\begin{eqnarray}
\widehat{\rm Ric}(X,Y) &=& {\rm Ric}(X,Y)-\tilde T(X,Y)\\
\widehat{\rm Ric}(X,\xi) &=& 0\\
\widehat{\rm Ric}(\xi,\xi) &=& -e^{2\theta}{\rm tr}(\tilde T)
\end{eqnarray}
\end{thm}
{\bf Proof:} Let $\{ E_1,\cdots E_n\}$ be an orthonormal local
base of $(M,g)$ and \mbox{$\hat i=<E_i,E_i>=\pm 1$}. Therefore,
$\{ E_1,\cdots E_n, e^{-\theta}\xi\}$ is an orthonormal local base
for $\hat TM$. Following computations show that (13) hold.
\begin{eqnarray*}
\widehat{\rm Ric}(X,Y) &=& \sum_{i=1}^n\hat i<\hat R(X,E_i)(E_i),Y>+<\hat R(X,e^{-\theta}\xi)(e^{-\theta}\xi),Y>\\
 &=&  \sum_{i=1}^n\hat i<R(X,E_i)(E_i),Y>+e^{-2\theta}<\hat R(X,\xi)(\xi),Y>\\
 &=& Ric(X,Y)-e^{-2\theta}<\hat R(X,\xi)(Y),\xi)>\\
 &=& Ric(X,Y)-e^{-2\theta}<\tilde T(X,Y)\xi,\xi>=Ric(X,Y)-e^{-2\theta}\tilde T(X,Y)<\xi,\xi>\\
 &=& {\rm Ric}(X,Y)-\tilde T(X,Y)
\end{eqnarray*}
Proof of (14):
\[
\widehat{\rm Ric}(X,\xi)= \sum_{i=1}^n\hat i<\hat
R(X,E_i)(E_i),\xi>+<\hat
R(X,e^{-\theta}\xi)(e^{-\theta}\xi),\xi>=0
\]
Proof of (15):
\begin{eqnarray*}
\widehat{\rm Ric}(\xi,\xi) &=& \sum_{i=1}^n\hat i<\hat R(\xi,E_i)(E_i),\xi>+<\hat R(\xi,e^{-\theta}\xi)(e^{-\theta}\xi),\xi>\\
 &=&  -\sum_{i=1}^n\hat i<\hat R(E_i,\xi)(E_i),\xi>= -\sum_{i=1}^n\hat i<\tilde T(E_i,E_i)\xi,\xi> \\
 &=& -\sum_{i=1}^n\hat i\tilde T(E_i,E_i)<\xi,\xi>=-e^{2\theta}{\rm tr}(\tilde T)\quad\bullet
\end{eqnarray*}
\begin{thm}
If scalar curvature of $\hat g$ and $g$ are denoted by $\hat R$
and $R$ respectively, then
\begin{equation} \hat R=R-2{\rm tr}(\tilde T) \end{equation}
\end{thm}
{\bf Proof:}
\[\begin{array}{rl} \hat R&=\sum_{i=1}^n\hat i\widehat {\rm Ric}(E_i,E_i)+\widehat{\rm Ric}(e^{-\theta}\xi,e^{-\theta}\xi)\\
 &=\sum_{i=1}^n\hat i\left( {\rm Ric}(E_i,E_i)-\tilde T(E_i,E_i)\right)+e^{-2\theta}\widehat{\rm Ric}(\xi,\xi)\\
 &=R-{\rm tr}(\tilde T)-{\rm tr}(\tilde T)=R-2{\rm tr}(\tilde T)\quad\bullet
\end{array}\]

\section{Application to General Relativity}
In this section we consider $M$ as a space-time manifold whose
dimension is $n$ and $2\leq n$. Assume $\hat g=(g,\theta)$ is a
graded metric on $\hat TM$. $g$ is an arbitrary semi-riemannian
metric in $M$ and play the role of potential for gravity. We will
find a field equation in which $\theta$ play the role of
potential for mass distribution in space-time. In this structure,
even part of $\hat TM$ is related to the structure of space and
time(gravity) and its odd part relate to the structure of mass
distribution. To find a proper field equation, we use Hilbert
action in the context of graded metrics on $\hat TM$. Denote
canonical volume form of a meter $g$ on oriented manifold $M$ by
$\Omega_g$. Hilbert action ${\cal L}$ on graded metrics is
defined as follows.
\[ {\cal L}(\hat g)={\cal L}(g,\theta)=\int_M\hat R\Omega_g \]
To be more precise, we must assume $M$ is compact or we must
integrate on open subset $U$ of $M$ such that $\overline U$ is
compact [1]. To find a critical meter $\hat g$ for Hilbert action
we must do some lengthy and tedious computations.

A variation for a meter $\hat g$ is obtained by a pair $(s,h)$ in
which $s$ is a symmetric 2-covariant tensor on $M$ and $h\in
C^\infty(M)$. Set $\tilde g(t)=g+ts$ and $\tilde\theta(t)=\theta
+th$. For small $t$, $\hat g(t)=(\tilde g(t),\tilde\theta(t))$ is
a graded metric and is a variation of $\hat g$. $\hat g$ is a
critical meter for Hilbert action iff there hold for every pair
$(s,h)$:
 \begin{equation}
 {\te d \over dt}|_{t=0}{\cal L}(\tilde g(t),\tilde\theta(t))= {\te d \over
dt}|_{t=0}\int_M\hat R(t)\Omega_{g+ts}=0
\end{equation}
$\hat R(t)$ is the scalar curvature of $\hat g(t)$ and
\[\hat R(t)=\tilde R(t)-2(\triangle^t(\tilde\theta(t))+|\vn^t\tilde\theta(t)|^2) \]
$\tilde R(t)$ is the scalar curvature of $\tilde g(t)$. Note that
in above, all Levi-Civita connection, gradient, divergence,
Laplacian, and volume form, depend on $t$. To find derivation in
(17), we must compute derivations of $\tilde R(t)$,
$\Omega_{g+ts}$, $\triangle^t(\tilde\theta(t))$,
$|\vn^t\tilde\theta(t)|^2$ for $t=0$.

Note that an inner product in a vector space, extend to an inner
product in tensor spaces. For example, if $T$ and $S$ are two
2-covariant tensors on an inner product space $V$, and $g_{ij}$
are components of the inner product in some bases of $V$, then
\[ <T,K>=g^{im}g^{jn}T_{ij}K_{mn}=T^{mn}K_{mn} \]
In particular, $<T,g>=g^{ij}T_{ij}={\rm tr}(T)$. Fix some chart
$(x,U)$ on $M$. For the sake of simplicity, denote
${\te\p\over\te\p x^i}$ by $\p_i$. Local components of the meter
$\tilde g(t)$ are denoted by $\tilde g_{ij}(t)$, so $\tilde
g_{ij}(t)=g_{ij}+ts_{ij}$ and $\tilde g'_{ij}(0)=s_{ij}$. By
$\tilde g^{ik}(t)\tilde g_{kj}(t)=\delta^i_j$ and derivation with
respect to $t$, it is deduced that \mbox{$(\tilde g^{ij})
'(0)=-s^{ij}$}.

Components of Levi-Civita connection of $\tilde g(t)$ are denoted
by $\tilde \Gamma_{ij}^k(t)$. It is well known that
\[\tilde \Gamma_{ij}^k(t)={\te 1\over 2}\tilde g^{kl}(t)
({\te\p\tilde g_{il}(t)\over\p x^j}+{\te\p\tilde g_{jl}(t)\over\p x^i}
-{\te\p\tilde g_{ij}(t)\over\p x^l})\]
 Derivation with respect to $t$, implies
\[
(\tilde \Gamma_{ij}^k)'(0)=-{\te 1\over 2}s^{kl}(t)({\te\p
g_{il}\over\p x^j}+{\te\p g_{jl}\over\p x^i}-{\te\p g_{ij}\over\p
x^l}) +{\te 1\over 2} g^{kl}({\te\p s_{il}\over\p x^j}+{\te\p
s_{jl}\over\p x^i}-{\te\p s_{ij}\over\p x^l})
\]
$(\tilde \Gamma_{ij}^k)'(0)$ is a tensor of type $(1,2)$, and we
denote it by $A$, so $A_{ij}^k= (\tilde \Gamma_{ij}^k)'(0)$.
Tensor $A$ is related to the 3-covariant tensor $\n s$ whose
components are $s_{ij,k}=(\n_{\p_k}s)(\p_i,\p_j)$. Direct
computations show that
\begin{equation}
 A^k_{ij}={\te 1\over 2} g^{kl}(s_{lj,i}+s_{li,j}-s_{ij,l})
\end{equation}
Denote components of the curvature tensor of $\tilde g(t)$ by
$(\tilde {\rm R}^l_{ijk})(t)$. Due to the formula of computation
of these components with respect to $\tilde \Gamma_{ij}^k(t)$ and
derivation with respect to $t$, we find that [1]
\begin{equation}
 (\tilde {\rm R}^l_{ijk})'(0)=A^l_{jk,i}-A^l_{ik,j}
\end{equation}
Denote Ricci curvature of $\tilde g(t)$ by $\widetilde{\rm
Ric}(t)$. Derivation of $\widetilde{\rm Ric}(t)$ for $t=0$ is as
follows.
\begin{equation}
\widetilde{\rm Ric}'_{jk}(0)= (\tilde
R^l_{ljk})'(0)=A^l_{jk,l}-A^l_{lk,j}
\end{equation}
Now we can find derivation of $\tilde R(t)$ at $t=0$.
\begin{eqnarray*}
 \tilde R'(0)&=&(\tilde g^{ij}(t)\widetilde{\rm Ric}_{ij}(t))'(0)=({\tilde g}^{ij})^\prime(0){\rm Ric}_{ij}+g^{ij}\widetilde{\rm Ric}'_{ij}(0)\\
&=&-s^{ij}{\rm Ric}_{ij}+g^{ij}\widetilde{\rm Ric}'_{ij}(0) =-<s\
,{\rm Ric}>+g^{ij}(A^l_{ij,l}-A^l_{lj,i})
\end{eqnarray*}
Define the vector field $W$ by components
$W^l=g^{ij}A_{ij}^l-g^{il}A_{ij}^j$, so \mbox{${\rm
div}(W)=g^{ij}(A^l_{ij,l}-A^l_{lj,i})$} [1]. Therefore,
\begin{equation}  \tilde R'(0)=-<s\ ,{\rm Ric}>+{\rm div}(W) \end{equation}
It is shown that the derivation of $\Omega_{g+ts}$ at $t=0$ is as
follows [1].
\begin{equation} \Omega'_{g+ts}(0)={\te 1\over 2}<g,s>\Omega_g \end{equation}
By local computations, we find derivation of
$\triangle^t(\tilde\theta(t))$ at $t=0$ . Local computation of
Laplacian of a smooth function $f$ is as follows.
\begin{equation} \triangle (f)=g^{ij}Hes(f)_{ij}=g^{ij}({\p^2f\over \p x^i\p x^j}-\Gamma_{ij}^k{\p f\over\p x^k}) \end{equation}
So,
\begin{eqnarray*}
\triangle^t(\tilde\theta(t))'(0)&=&\triangle^t(\theta
+th)'(0)=\left( \tilde g^{ij}(t)({\p^2(\theta+th)\over \p x^i\p
x^j}-\tilde\Gamma_{ij}^k(t) {\p (\theta+th)\over\p
x^k})\right)'(0)\\
 &=&-s^{ij}({\p^2\theta\over \p x^i\p x^j}-\Gamma_{ij}^k{\p \theta\over\p x^k})+g^{ij}({\p^2h\over \p x^i\p x^j}-A_{ij}^k{\p\theta\over\p x^k}
-\Gamma_{ij}^k{\p h\over\p x^k})\\
 &=& -<s,{\rm Hes}(\theta)>+\triangle(h)-g^{ij}A_{ij}^k{\p\theta\over\p
 x^k}\qquad\qquad\qquad\qquad (24)
\end{eqnarray*}
 Define the vector field $Y$ by components
$Y^k=g^{ij}A_{ij}^k$. So,
\begin{eqnarray*}
Y^k&=&g^{ij}A_{ij}^k={\te 1\over 2}g^{ij} g^{kl}(s_{lj,i}+s_{li,j}-s_{ij,l})={\te 1\over 2} g^{kl}(g^{ij}s_{lj,i}+g^{ij}s_{li,j}-g^{ij}s_{ij,l})\\
 &=& {\te 1\over 2} g^{kl}({\rm div}(s)_l+{\rm div}(s)_l-{\rm tr}(s)_{,l})={\rm div}(s)^k-{1\over 2}(\vn {\rm tr}(s))^k
\end{eqnarray*}
Consequently,
\[ g^{ij}A_{ij}^k{\p\theta\over\p
x^k}={\rm div}(s)^k{\p\theta\over\p x^k}-{1\over 2}(\vn {\rm
tr}(s))^k{\p\theta\over\p x^k} ={\rm div}(s)(\vn\theta)-{1\over
2}<\vn {\rm tr}(s)\ ,\vn\theta > \] In the following, whenever it
is convenient, we consider $s$ as a (1,1) symmetric tensor. For
arbitrary vector field $Z$ we have
\begin{eqnarray*}
{\rm div}(s)(Z)&=&g^{ij}<(\n_{\p_i}s)(Z)\ ,\p_j>=g^{ij}<\n_{\p_i}s(Z)-s(\n_{\p_i}Z)\ ,\p_j>\\
 &=& g^{ij}<\n_{\p_i}s(Z)\ ,\p_j>-g^{ij}<s(\n_{\p_i}Z)\ ,\p_j>\\
 &=&{\rm div}(s(Z))-g^{ij}<s(\p_j)\ ,\n_{\p_i}Z>={\rm div}(s(Z))-<s\ ,\n Z>
\end{eqnarray*}
For $Z=\vn\theta$, the $(1,1)$ tensor $\n\vn\theta$ as a
2-covariant tensor is $\n d\theta$ and is equal to ${\rm
Hes}(\theta)$. So,
\[  g^{ij}A_{ij}^k{\p\theta\over\p x^k}={\rm div}(s(\vn\theta))-<s\ ,{\rm Hes}(\theta)>-{1\over 2}<\vn {\rm tr}(s)\ ,\vn\theta > \]
Now, we go back to computations in (24). By above computations, we
have
\begin{eqnarray*}
\triangle^t(\tilde\theta(t))'(0)&=& -<s,{\rm
Hes}(\theta)>+\triangle(h)-g^{ij}A_{ij}^k{\p\theta\over\p
x^k}\\
 &=&-<s,{\rm Hes}(\theta)>+\triangle(h)\\
 &&-({\rm div}(s(\vn\theta))-<s\ ,{\rm Hes}(\theta)>-{1\over 2}<\vn {\rm tr}(s)\ ,\vn\theta >)\\
 &=&\triangle(h)-{\rm div}(s(\vn\theta))+{1\over 2}<\vn {\rm tr} (s)\ ,\vn\theta >\qquad\qquad\qquad (25)
\end{eqnarray*}
Now, we compute derivation of $|\vn^t(\tilde\theta(t))|^2$ at
$t=0$.
\begin{eqnarray*}
{d\over dt}|_{t=0}|\vn^t(\tilde\theta(t))|^2 &=&( \tilde
g^{ij}(t){\p\tilde\theta(t)\over\p x^i}{\p\tilde\theta(t)\over\p
x^j})'(0) =-s^{ij}{\p\theta\over\p x^i}{\p\theta\over\p
x^j}+2g^{ij}{\p h\over\p x^i}{\p\theta\over\p x^j}\\
&=& -<s\ ,d\theta\otimes d\theta>+2<\vn h\
,\vn\theta>\qquad\qquad\qquad (26)
\end{eqnarray*}
\setcounter{equation}{26} Remind that the integral of divergence
of every vector fields on $M$ is zero, consequently integral of
Laplacian of any smooth function on $M$ is zero. Also, for any
two smooth function $f$ and $h$ we have
\[ \int_M<\vn h\ ,\vn f>\Omega_g=-\int_Mh\triangle(f)\Omega_g \]
Now, we are ready to find critical metric $\hat g$ for Hilbert
action.
\begin{eqnarray*}
{\cal L}(\tilde g(t),\tilde\theta(t))'(0)&=& \left(\int_M\hat
R(t)\Omega_{g+ts}\right)'(0)
= \int_M\left( (\tilde R(t)-2(\triangle^t(\tilde\theta(t))+|\vn^t\tilde\theta(t)|^2)\Omega_{g+th}\right)'(0)\\
&=& \int_M(R-2\triangle(\theta)-2|\vn\theta|^2)\Omega'_{g+ts}(0)\\
&&+(\tilde R'(0)-2\triangle^t(\tilde\theta(t))'(0)-2(|\vn^t(\tilde\theta(t))|^2)'(0)\Omega_g\\
&=&\int_M(R-2\triangle(\theta)-2|\vn\theta|^2)\times{1\over 2}<s\ ,g>\Omega_g+\left(-<s\ ,{\rm Ric}>+{\rm div}(W)\right.\\
&& -2(\triangle(h)-{\rm div}(s(\vn\theta))+{1\over 2}<\vn {\rm tr}(s)\ ,\vn\theta>)\\
&&-2(-<s\ ,d\theta\otimes d\theta>+2<\vn h\ ,\vn\theta>))\Omega_g\\
&=&\int_M ( <s\ ,({1\over 2}R-\triangle\theta-|\vn\theta|^2)g>-<s\ ,Ric> \\
&&+{\rm tr}(s)\triangle(\theta)+2<s\ ,d\theta\otimes d\theta>+4h\triangle(\theta))\Omega_g\\
&=& \int_M \left(<s\ ,-{\rm Ric}+({1\over
2}R-|\vn\theta|^2)g+2d\theta\otimes
d\theta>+4h\triangle(\theta)\right)\Omega_g
\end{eqnarray*}
The above expression is zero for all pair $(s,h)$ iff
\begin{eqnarray}
{\rm Ric}-{1\over 2}Rg&=&2d\theta\otimes d\theta-|\vn\theta|^2g\\
\triangle(\theta)&=&0
\end{eqnarray}
These are field equations for $\hat g=(g,\theta)$ and determine
critical metrics for Hilbert action. Equation (28) express that
${\rm div}(\vn \theta)=0$, so $\vn\theta$ can be interpreted as
current of mass that satisfies conservation law. Absolute value of
$2|\vn\theta|^2$ shows density of matter. Right hand side of
equation (27) can be interpreted as energy-momentum tensor of
matter and the following theorem shows that it satisfies
conservation law.
\begin{thm}
If $\theta$ be a smooth function on a semi-riemannian manifold
$(M,g)$ such that $\triangle(\theta)=0$, then the divergence of
symmetric tensor $2d\theta\otimes d\theta-|\vn\theta|^2g$ is zero.
\end{thm}
{\bf Proof:} Let $\{ E_1,\dots,E_n\}$ be an orthonormal local
base on $M$ and \mbox{$\hat i=<E_i\ ,E_i>=\pm 1$}. So,
\begin{eqnarray*}
{\rm div}(d\theta\otimes d\theta)(X)&=&\sum_{i=1}^n\hat
i(\nabla_{E_i}d\theta\otimes d\theta)(E_i\ ,X)\\
&=&\sum_{i=1}^n\hat i((\n_{E_i}d\theta)\otimes d\theta+d\theta\otimes(\n_{E_i}d\theta))(E_i\ ,X)\\
 &=&\sum_{i=1}^n\hat i \left( (\n_{E_i}d\theta)(E_i)d\theta(X)+d\theta(E_i)(\n_{E_i}d\theta)(X)\right)\\
&=&\triangle(\theta)d\theta(X)+\sum_{i=1}^n\hat i
d\theta(E_i){\rm Hes}(\theta)(E_i,X)={\rm
Hes}(\theta)(\vn\theta,X)
\end{eqnarray*}
Remind that for any smooth function $h$ on $M$: ${\rm
div}(hg)=dh$. So,
\begin{eqnarray*}
{\rm div}(|\vn\theta|^2g)(X)&=&d(|\vn\theta|^2)(X)=X<\vn\theta\ ,\vn\theta>=2<\nabla_X(\vn\theta)\ ,\vn\theta>\\
&=&2(\n_Xd\theta)(\vn\theta)=2Hes(\vn\theta\ ,X)
\end{eqnarray*}
The above computations show that ${\rm div}(2d\theta\otimes
d\theta-|\vn\theta|^2g)=0.\quad\bullet$

In case $3\leq {\rm dim}(M)$, equation (27) can be written in a
simple form.
\begin{thm}
Suppose $3\leq n={\rm dim}(M)$. The metric $\hat g=(g,\theta)$ on
$\hat TM$, is a critical metric for Hilbert action iff
\begin{eqnarray}
Ric&=&2d\theta\otimes d\theta \\
\triangle(\theta)&=&0
\end{eqnarray}
\end{thm}
{\bf Proof:} Equation (30) is the same as (28). First suppose
equation (27) holds. Compute traces of tow sides of equation (27).
\[R-{n\over 2}R=2|\vn\theta|^2-n|\vn\theta|^2\ \Rightarrow\ {2-n\over 2}R=(2-n)|\vn\theta|^2\ \Rightarrow\ R=2|\vn\theta|^2\]
By substitution into (27) we obtain (29). Now, assume (29) holds.
By computation traces of tow sides of equation (29), we find
$R=2|\vn\theta|^2$, so by addition suitable expression to each
side of equation (29) we obtain (27).$\quad\bullet$

{\bf Example:} Suppose $(N,\bar g)$ be a Riemannian manifold of
dimension $n\geq 2$. Set $M=N\times (0,\infty)$. Tangent vectors
to $M$ at $(p,t)$ is of the form $(v,\lambda)$ in which $v\in
T_pN$ and $\lambda\in\R$. Denote the vector field $(0,1)$ on $M$
by $\p_t$. As a derivation, for a smooth function $f(p,t)$ on $M$
we have $\p_t(f)={\te\p f\over\te\p t}$. Denote vector fields on
$N$ by $X$, $Y$, $Z$, ... and consider them as special vector
fields on $M$. Denote second projection map $(p,t)\mapsto t$ on
$M$ by $t$. we can interpret $t$ as time. For the 1-form $dt$ we
have $dt(v,\lambda)=\lambda$, so $dt(\p_t)=1$. For some smooth
function $a:(0,\infty)\lon\R$ define a metric $g$ on $M$ as
follows.
\[ g=e^{2a(t)}\bar g-dt\otimes dt \]
So, inner product of special vector fields on $M$ and $\p_t$ are
as follows.
\[ <X,Y>=e^{2a}\bar g(X,Y)\quad,\quad <X,\p_t>=0\quad,\quad <\p_t,\p_t>=-1 \]
Consider $\theta:N\times(0,\infty)\lon\R$ such that its level
sets be $N\times\{ t\}$. Therefore, $\theta(p,t)$ must depend only
on $t$, and we denote it by $\theta(t)$. Consequently,
$d\theta=\theta'(t)dt$ and $\vn\theta=-\theta'(t)\p_t$ and
$|\vn\theta|^2=<\vn\theta,\vn\theta>=-|\theta'(t)|^2$.

Denote Levi-Civita connection of $N$ and $M$ by $\bn,\ \n$
respectively. Straightforward computations show that:
\begin{eqnarray}
\n_XY&=&\bn_XY+a'<X,Y>\p_t\\
\n_{\p_t}Y&=&a'Y\\
\n_{\p_t}\p_t&=&0
\end{eqnarray}
Suppose $\{ E_1,\cdots E_n\}$ is an orthonormal local base on $N$,
then, $\{e^{-a}E_1\cdots,e^{-a}E_n,\p_t\}$ is an orthonormal local
base on $M$. Laplacian of $\theta$ can be computed as follows.
\begin{eqnarray*}
\triangle\theta&=&\sum_{i=1}^ne^{-2a}<\n_{E_i}\vn\theta,E_i>-<\n_{\p_t}\vn\theta,\p_t>\\
&=& \sum_{i=1}^ne^{-2a}<\n_{E_i}(-\theta'(t)\p_t),E_i>-<\n_{\p_t}(-\theta'(t)\p_t),\p_t>\\
&=& \sum_{i=1}^n-\theta'(t)e^{-2a}<a'E_i,E_i>-\theta''(t)\\
&=&
(\sum_{i=1}^n-a'\theta'(t))-\theta''(t)=-na'\theta'(t)-\theta''(t)
\end{eqnarray*}
Equation $\triangle\theta=0$, implies that for some constant $c$
we have $|\theta'(t)|=ce^{-na(t)}$.

Denote curvature tensors and Ricci curvature tensors of $N$ and
$M$ respectively by $\bar {\rm R},\ {\rm R},\ \overline{\rm
Ric},\ {\rm Ric}$. Straight forward computations show that:
\begin{eqnarray}
{\rm R}(X,Y)(Z)&=&\bar {\rm R}(X,Y)(Z)+a'^2(<Y,Z>X-<X,Z>Y)\\
{\rm R}(X,Y)(\p_t)&=&0\\
{\rm R}(\p_t,Y)(Z)&=&(a''+a'^2)<Y,Z>\p_t\\
{\rm R}(\p_t,Y)(\p_t)&=&(a''+a'^2)Y\\
{\rm Ric}(X,Y)&=&\overline{\rm Ric}(X,Y)+(a''+na'^2)<X,Y>\\
{\rm Ric}(X,\p_t)&=&0\\
{\rm Ric}(\p_t,\p_t)&=&-n(a''+a'^2)
\end{eqnarray}
In this example, equation (29) becomes ${\rm
Ric}=2d\theta\otimes\theta=2|\theta'(t)|^2dt\otimes dt$, ant it
holds iff
\begin{eqnarray}
\overline{\rm Ric}(X,Y)&=&-(a''+na'^2)<X,Y>\\
-n(a''+a'^2)&=&2|\theta'(t)|^2=2c^2e^{-2na(t)}
\end{eqnarray}
Left side of (41) dose not depend on $t$, so $a''+na'^2$ must be
constant and $N$ is an Einstein manifold. In the case
$a''+na'^2=0$ we find solution $a(t)={1\over n}{\rm ln}( t)$ and
for this solution, equation (42) also holds for $c=\sqrt{\te
n-1\over \te 2n}$.

So, for an $n$ dimensional Ricci flat manifold $(N,\bar g)$ the
meter $g=t^{2\over n}\bar g-dt\otimes dt$ on $M$ and the function
$\theta(t)= \sqrt{\te n-1\over \te 2n}{\rm ln}(t)$ satisfies
field equations (29) and (30). In this model, as $t$ approaches
zero, universe become smaller and density of matter increases to
infinity. Time $t=0$ is not in $M$ and this time is the instant of
Big-Bang. This example is an Einstein-de Sitter model in general
relativity.

\section{Field equation on the graded tangent bundle}
We have found two equations (29) and (30) as field equations on
$M$. But, we can incorporate these equations to one equation on
$\hat TM$.

Exterior derivation in algebroid structures is defined.
Specially, for a smooth function $f$, its exterior derivation,
denoted by $\hat df$, is a 1-form on $\hat TM$ defined by $\hat
df(\hat X)=\hat X(f)$. Restriction of $\hat df$ to $TM$ is $df$,
and its restriction to odd vectors is zero. Hessian of $f$ which
is denoted by $\widehat{\rm Hes}(f)$, is a 2-covariant symmetric
tensor on $\hat TM$ and defined as follows.
\[ \widehat{\rm Hes}(f)(\hat X\ ,\hat Y)=(\hn_{\hat X}\hat df)(\hat Y)\qquad \hat X,\hat Y\in \hat \X (M) \]
Straight computations show that $\widehat{\rm Hes}(f)$ satisfies
the following relations.
\begin{eqnarray*}
\widehat{\rm Hes}(f)(X,Y)&=& {\rm Hes}(f)(X,Y)\\
\widehat{\rm Hes}(f)(\xi,X)&=&0\\
\widehat{\rm Hes}(f)(\xi,\xi)&=&e^{2\theta}<\vn f,\vn\theta>
\end{eqnarray*}
Consequently, ${\rm tr}(\widehat{\rm Hes}(f))=\triangle (f)+<\vn
f,\vn\theta>$. In particular,
\begin{equation} {\rm tr}(\widehat{\rm Hes}(\theta))=\triangle(\theta)+|\vn\theta|^2={\rm tr}(\tilde T) \end{equation}
\begin{thm}
Let $M$ be a manifold, then a pair $(g,\theta)$ satisfies in field
equations (29) and (30) iff for the meter $\hat g=(g,\theta)$:
\begin{equation}  \widehat{\rm Ric}=\hat d\theta\otimes\hat d\theta-\widehat{\rm Hes}(\theta) \end{equation}
$\widehat{\rm Ric}$ is the Ricci curvature of the meter $\hat g$.
\end{thm}
{\bf Proof:} First, suppose (44) holds. So,
\[ \begin{array}{rl}
 &\widehat{\rm Ric}(\xi,\xi)=(\hat d\theta\otimes\hat d\theta-\widehat{\rm Hes}(\theta))(\xi,\xi)=-\widehat{\rm Hes}(\theta)(\xi,\xi)\\
\Rightarrow & -e^{2\theta}{\rm tr}(\tilde T)=-e^{2\theta}|\vn\theta|^2\ \Rightarrow\ {\rm tr}(\tilde T)=|\vn\theta|^2   \\
\Rightarrow &\triangle(\theta)+|\vn\theta|^2=|\vn\theta|^2 \ \
\Rightarrow\ \ \triangle(\theta)=0
\end{array}\]
Therefore, (30) holds. Moreover,
\[ \begin{array}{rl}
 &\widehat{\rm Ric}(X,Y)=(\hat d\theta\otimes\hat d\theta-\widehat{\rm Hes}(\theta))(X,Y)\\
\Rightarrow & ({\rm Ric}-\tilde T)(X,Y)= (d\theta\otimes d\theta-{\rm Hes}(\theta))(X,Y)\\
\Rightarrow &  {\rm Ric} -{\rm Hes}(\theta)-d\theta\otimes d\theta=d\theta\otimes d\theta-{\rm Hes}(\theta)\\
\Rightarrow & {\rm Ric}=2d\theta\otimes d\theta
\end{array} \]
Therefore, (29) holds. Conversely, suppose (29) and (30) hold.
All above computations are reversible, and show that each side of
(44) are equal on even and odd vectors. Also, we can see
directly, that each side of (44) on an even and an odd vectors are
zero. So, (44) holds. $\quad\bullet$

 \end{document}